\documentclass[twocolumn,showpacs,showkeys]{revtex4}
\usepackage{graphicx}
\input{psfig.sty}

\newcommand{\bastar}{\begin{eqnarray*}}
\newcommand{\eastar}{\end{eqnarray*}}
\newskip\humongous \humongous=0pt plus 1000pt minus 1000pt

\newif\ifdtup

\relax
\newcommand{\be}{\begin{equation}}
\newcommand{\ee}{\end{equation}}
\newcommand{\bea}{\begin{eqnarray}}
\newcommand{\eea}{\end{eqnarray}}
\newcommand{\X}{{\vec X}}
\newcommand{\pro}{\partial}
\newcommand{\n}{\hat n}
\newcommand{\oneg}{\displaystyle\frac{1}{g}}

\newcommand{\D}{{\hat D}}

\newcommand{\A}{{\vec A}}
\newcommand{\valpha}{{\vec \alpha}}

\newcommand{\dfrac}{\displaystyle\frac}
\newcommand{\ba}{\begin{array}}
\newcommand{\ea}{\end{array}}

\newcommand{\nn}{\nonumber}
\newcommand{\hn}{\hat n}
\begin{document}
\title{Non-Abelian Superconductivity}
\bigskip
\author{Y. M. Cho}
\email{ymcho@yongmin.snu.ac.kr}
%\author{}
%\email{}
\affiliation{School of Physics, College of Natural Sciences,
Seoul National University,
Seoul 151-742, Korea  \\
and \\
C.N.Yang Institute for Theoretical Physics, State University of
New York, Stony Brook, NY 11790, USA}
\begin{abstract}
~~~We establish a non-Abelian superconductivity and
a non-Abelian Meissner effect by constructing
an effective field theory of superconductivity 
in which a genuine $SU(2)$ gauge symmetry governs the
dynamics. We show that the magnetic flux is quantized in the unit 
of $4\pi/g$, not $2\pi/g$, in the non-Abelian superconductor.
\end{abstract}
\maketitle

The superconductivity has always been based on quantum 
electrodynamics which is an Abelian gauge theory. The only 
exception is the $SO(5)$ model of high-{$\rm T_c$} 
superconductivity \cite{zhang}. In this view 
one might assume that the underlying dynamics of the recently 
discovered two-gap superconductor made of ${\rm MgB_2}$ \cite{sc} 
should also be the Abelian (i.e., electromagnetic) interaction. 
An important feature of multi-gap supercondectors,
however, is its non-Abelian structure,
since multi-gap supercondectors can be described 
by multi-component condensates which can naturally form 
a non-Abelian multiplet \cite{cm2}. This raises 
the possibility of a non-Abelian superconductivity
and a non-Abelian Meissner effect.

The motivation for a non-Abelian superconductivity must be clear.
Suppose the underlying dynamics of the two-gap superconductor 
is the Abelian interaction. If so,
the effective theory of two-gap superconductor
should be an Abelian gauge theory which has a global
$SU(2)$ symmetry. But in this case
the two condensates must carry the same charge 
(two electron-electron pairs or two hole-hole pairs), because
there is no way that the Abelian gauge field can
couple to a doublet condensate whose components
have opposite charges (one electron-electron pair and
one hole-hole pair). This means that a two-gap superconductor
made of a doublet whose components have opposite charges
can not be described by an Abelian gauge theory
(unless the two components do not form a doublet).
This poses a problem because this
implies that one can not construct a theory of
two-gap superconductor made of oppositely charged condensates
based on an Abelian gauge theory.

{\it The purpose of this paper is to establish
a non-Abelian superconductivity and a non-Abelian Meissner 
effect by constructing a non-Abelian
gauge theory of superconductivity in which the magnetic flux 
is quantized in the unit $4\pi/g$.} 
We present an $SU(2)$ gauge theory of 
superconductivity which is governed by a genuine 
non-Abelian dynamics, which can describe a two-gap superconductor 
made of an oppositely charged doublet condensate.
As far as we understand, this type of 
non-Abelian superconductivity
has never been discussed before.

To understand the non-Abelian superconductivity we must understand 
the Abelian two-gap superconductor first, because
the non-Abelian superconductor is closely related to 
the Abelian two-gap superconductor. So we start with
the Abelian two-gap superconductor and 
establish the Meissner effect in the theory first.
 
Consider a charged doublet scalar field $\phi$ coupled to
the electromagnetic field
\bea
&{\cal L} = - |D_\mu \phi|^2 + \mu^2 \phi^{\dagger}\phi
- \dfrac{\lambda}{2} (\phi^{\dagger} \phi)^2
- \dfrac{1}{4} F_{\mu \nu}^2, 
\label{sclag}
\eea
where $D_\mu \phi = (\partial_\mu + ig A_\mu) \phi$. 
This is an obvious generalization of the Landau-Ginzburg
Lagrangian. The Lagrangian has the equation of motion
\bea
&D^2\phi =\lambda(\phi^{\dagger} \phi
-\dfrac{\mu^2}{\lambda})\phi, \nn\\
&\partial_\mu F_{\mu \nu} = j_\nu =  i g \Big[(D_\nu
\phi)^{\dagger}\phi - \phi ^{\dagger}(D_\nu \phi) \Big],
\label{sceq1}
\eea
which tells that a non-vanishing $<\phi^{\dagger}\phi>$ 
makes the photon massive. This implies the existence of 
Meissner effect. To demonstrate the Meissner effect 
we construct a magnetic vortex in this theory.
Let 
\bea
&\phi =\dfrac{1}{\sqrt 2} \rho \xi,
~~~~~{\xi}^{\dagger}\xi = 1, \nn\\
&\rho=\rho(\varrho),  
~~~~~\xi = \Bigg( \matrix{\cos \dfrac{f(\varrho)}{2} \exp (-im\varphi) \cr
\sin \dfrac{f(\varrho)}{2} } \Bigg), \nn\\
&A_\mu= \dfrac{m}{g} A(\varrho) \partial_\mu \varphi,
\label{scans}
\eea
and find that (\ref{sceq1}) is reduced to
\bea
&\ddot{\rho}+\dfrac{1}{\varrho}\dot\rho
- \Big[\dfrac{1}{4}\Big(\dot{f}^2
+\dfrac{m^2}{\varrho^2}\sin^2{f}\Big) \nn\\
&+ \dfrac{m^2}{\varrho^2} \Big(A-\dfrac{\cos{f}+1}{2}\Big)^2\Big]\rho 
= \dfrac{\lambda}{2}(\rho^2-\rho_0^2)\rho, \nn\\
&\ddot{f} + \Big(\dfrac{1}{\varrho}+2\dfrac{\dot{\rho}}{\rho} \Big)\dot{f}
- 2\dfrac{m^2}{\varrho^2} \Big(A-\dfrac{1}{2} \Big) \sin{f} =0, \nn\\
&\ddot{A}-\dfrac{\dot{A}}{\varrho} -g^2 \rho^2
\Big(A-\dfrac{\cos{f}+1}{2}\Big) = 0.
\label{sceq4}
\eea
Now, we impose the following boundary condition for the non-Abelian
vortices \cite{cm2},
\bea
&\rho (0) = 0,~~~\rho(\infty) = \rho_0,
~~~f (0) = \pi,~~~f (\infty) = 0, \nn\\
& A (0) = -1,~~~A (\infty) = 1.
\label{scbc}
\eea
This need some explanation, because the boundary
value $A(0)$ is chosen to be $-1$, not $0$. This is to
assure the smoothness of the scalar field $\rho(\varrho)$
at the origin. Only with this boundary value $\rho$
becomes analythic at the origin.
One might object the boundary condition, because it creates
an apparent singularity in the gauge potential at the origin.
But notice that this singularity is an unphysical
(coordinate) singularity which can easily be removed by 
a gauge transformation.
In fact the singularity disappears with the gauge transformation
\bea
A_\mu \rightarrow A_\mu + \dfrac{m}{g} \partial_\mu \varphi,
\eea
which simultaneously changes the boundary condition $A(0)=-1,~A(\infty)=1$
to $A(0)=0,~A(\infty)=2$. 

\begin{figure}
    \includegraphics[scale=0.7]{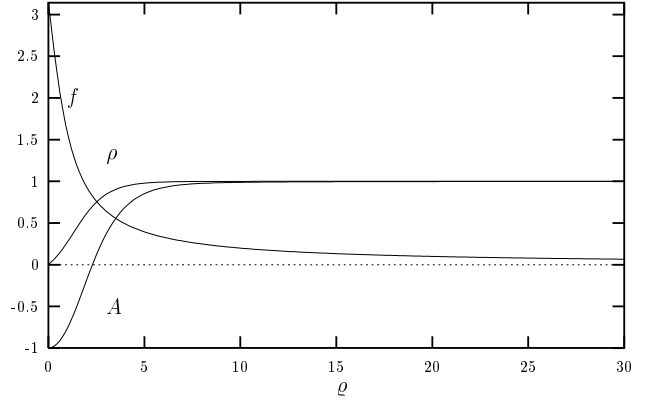}
    \caption{The non-Abelian vortex with $m=1$ in two-gap superconductor.
Here we have put $g=\lambda=1$, and $\varrho$ is in the unit of $\rho_0$.}
    \label{scv}
\end{figure}

With the boundary condition we can integrate (\ref{sceq4})
and obtain the non-Abelian vortex solution
of the two-gap superconductor, which is shown in Fig.\ref{scv}.
Notice that the boundary condition (\ref{scbc}) assures that 
the doublet $\xi$ starts from the second component at the origin 
and ends up with the first component at the infinity,
which shows that the non-Abelian vortex is different from
the well-known Abrikosov vortex in ordinary (one-gap)
superconductor \cite{abri}.

Clearly the magnetic field $H$ of the vortex
has total flux given by
\bea
&\hat \phi = \dfrac{}{} \int H d^2x
= \dfrac{2\pi m}{g} \big[A(\infty) - A(0)\big] 
= \dfrac{4\pi m}{g}.
\label{scflux}
\eea
Notice that the unit of the non-Abelian flux is $4\pi/g$,
not $2\pi/g$. This is the non-Abelian quantization of magnetic 
flux that we have in two-gap superconductor. 
We emphasize that this is a direct
consequence of the boundary condition
$A(0)=-1$ (or more precisely $A(\infty)-A(0)=2$)
in (\ref{scbc}). 

But just as in ordinary superconductor 
this non-Abelian quantization of magnetic flux
is of topological origin. To understand this notice that the doublet
$\xi$, with the $U(1)$ gauge symmetry, forms a $CP^1$ field.
So it can naturally define a mapping $\pi_2(S^2)$ from
the compactified $xy$-plane $S^2$ to the target space $S^2$,
which can be classified by the flux quantum number
\bea
&q = - \dfrac {1}{4\pi} \int \epsilon_{ij} \partial_i \xi^{\dagger}
\partial_j \xi d^2 x = m.
\label{scqn}
\eea
Again this non-Abelian topology should be compared with
the well-known Abelian topology $\pi_1(S^1)$ of the Abrikosov vortex.

The existence of the magnetic vortex solution demonstrates
that the Lagrangian (\ref{sclag}) generates the Meissner effect,
and thus can describe the superconductivity in two-gap
superconductor \cite{cm2}.

With the ansatz (\ref{scans}) one can express the Hamiltonian
of the vortex as
\bea
&{\cal H} = \dfrac{1}{2} \Big[ \big(\dot \rho
\pm \dfrac{m}{\varrho}(A-\dfrac{\cos f +1}{2}) \rho \big)^2 \nn\\
&+(\dot f \pm \dfrac{m}{\varrho} \sin f)^2 \dfrac{\rho^2}{4}
+ \big(H \pm \dfrac{\sqrt \lambda}{2}(\rho^2-\rho_0^2)\big)^2 \nn\\
&\pm H \big( g\rho^2 + \sqrt \lambda (\rho_0^2-\rho^2) \big) \Big],
\label{scham}
\eea
so that the Hamiltonian has a minimum value when
\bea
&\dot \rho \pm \dfrac{m}{\varrho}(A-\dfrac{\cos f +1}{2}) \rho =0, \nn\\
&\dot f \pm \dfrac{m}{\varrho} \sin f =0, \nn\\
&H \pm \dfrac{\sqrt \lambda}{2}(\rho^2-\rho_0^2) =0.
\label{scfoeq}
\eea
This can be viewed as a first order equation for the magnetic
vortex. Indeed, when the coupling constant $\lambda$ has 
the critical value (i.e., when $\lambda=g^2$), one can integrate 
and reduce the second order equation (\ref{sceq4}) to
the above first order equation.

Integrating the second equation of (\ref{scfoeq}) we have
\bea
\cos f(\varrho) = \dfrac{\varrho^{2m} - a^2}{\varrho^{2m} + a^2},
~~~\sin f(\varrho) = \dfrac{2a\varrho^m}{\varrho^{2m} + a^2},
\eea
where $a$ is an integration constant,
so that (\ref{scfoeq}) is reduced to (with $\lambda=g^2$)
\bea
&\dot \rho \pm \dfrac{m}{\varrho}(A
-\dfrac{\varrho^{2m}}{\varrho^{2m} + a^2}) \rho =0,  \nn\\
&\dfrac{m}{g} \dfrac{\dot A}{\varrho} \pm \dfrac{g}{2}(\rho^2-\rho_0^2) =0.
\eea
In this case the Hamiltonian becomes
\bea
{\cal H} = \dfrac{g}{2} H \rho_0^2,
\eea
and the energy (per unit length) acquires
the absolute minimum value
\bea
E = 2\pi m \rho_0^2 = \dfrac{g}{2} \rho_0^2 \hat \phi.
\eea
This tells that the minimum energy is fixed by the topological
flux quantum mumber.

This means that the non-Abelian vortex has twice as much
magnetic flux and energy. Again this difference can be traced back 
to the boundary conditions (\ref{scbc}).
Mathematically this difference
has a deep reason, which originates from the fact
that the Abelian $U(1)$ runs from $0$ to $2\pi$, 
but the $S^1$ fiber of $SU(2)$ runs from $0$ to $4\pi$ \cite{cm2}.

With the above preliminaries, we now construct 
an effective theory of two-gap superconductor which is based on
a $SU(2)$ gauge theory.
To do this we need to understand
the mathematical structure of the $SU(2)$ gauge theory better.
A best way to do this is to start with the well-known 
decomposition of the gauge potential \cite{cho80,fadd}. 
Let $\hn$ be the gauge covariant unit triplet
which select the charge direction in $SU(2)$, and decompose the
non-Abelian gauge potential into the restricted potential $\hat A_\mu$
and the valence potential $\X_\mu$,
\bea
& \vec{A}_\mu =A_\mu \n -
\oneg \n\times\pro_\mu\n+\X_\mu\nonumber
         = \hat A_\mu + \X_\mu, \nn\\
&  (A_\mu = \n\cdot \vec A_\mu,~ \n^2 =1,~
\hat{n}\cdot\vec{X}_\mu=0),
\label{cdecom}
\eea
where $ A_\mu$ is the
``electric'' potential. Notice that the restricted potential
is precisely the potential which leaves $\n$
invariant under the parallel transport,
\bea
\D_\mu \n = \pro_\mu \n
+ g {\hat A}_\mu \times \n = 0. 
\eea 
Under the infinitesimal gauge transformation 
\bea 
\delta \n = - \vec \alpha \times \n\,,\,\,\,\, 
\delta \A_\mu = \oneg  D_\mu \vec \alpha, 
\eea 
one has
\bea 
&\delta A_\mu = \oneg \n \cdot \pro_\mu \valpha,\,\,\,\
\delta \hat A_\mu = \oneg \D_\mu \valpha,  \nn \\
&\delta \X_\mu = - \valpha \times \X_\mu.
\eea
This tells that restricted potential $\hat A_\mu$ 
by itself describes an $SU(2)$
connection which enjoys the full $SU(2)$ gauge degrees of
freedom. Furthermore the valence potential $\vec X_\mu$ forms a
gauge covariant vector field under the gauge transformation.
More importantly, the decomposition is
gauge-independent. Once the gauge covariant topological field
$\hat n$ is given, the decomposition follows automatically
independent of the choice of a gauge \cite{cho80}.

The importance of the decomposition (\ref{cdecom}) for our
purpose is that we can construct a non-Abelian gauge theory
which has a full non-Abelian gauge
degrees of freedom, with the restricted potential
$\hat A_\mu$ alone \cite{cho80}. This is because
the valence potential $\vec X_\mu$ can be treated as
a gauge covariant source, so that one can always exclude it from
the theory without compromizing the gauge invariance.
Indeed we will see that it is this
restricted gauge theory which describes the non-Abelian 
superconductivity.

To demonstrate the non-Abelian superconductivity consider the
Lagrangian which is made of two condensates which couple
to the restricted $SU(2)$ gauge field,
\bea
{\cal L} = -|\hat D_\mu \Phi|^2 + \mu^2 \Phi^{\dagger}\Phi
- \dfrac{\lambda}{2} (\Phi^{\dagger} \Phi)^2
-\dfrac{1}{4} {\hat F}_{\mu\nu}^2,
\label{nasclag1}
\eea
where now
\bea
\hat D_\mu \Phi = ( \partial_\mu + \dfrac{g}{2i} \vec \sigma
\cdot \hat A_\mu ) \Phi. \nn
\eea
The equation of
motion of the Lagrangian is given by
\bea
&{\hat D}^2\Phi
=\lambda(\Phi^{\dagger} \Phi
-\dfrac{\mu^2}{\lambda})\Phi, \nn\\
&\hat D_\mu \hat F_{\mu \nu} = \vec j_\nu
= g \Big[(\hat D_\nu \Phi)^{\dagger}\dfrac{\vec\sigma}{2i} \Phi
- \Phi ^{\dagger} \dfrac{ \vec\sigma}{2i}(\hat D_\nu \Phi) \Big].
\label{nasceq}
\eea
Let $\xi$ and $\eta$
be two orthonormal doublets which form a basis,
\bea
&\xi^{\dagger}\xi =1,~~~~~\eta^{\dagger} \eta = 1,
~~~~~\xi^{\dagger}\eta = \eta^{\dagger} \xi = 0, \nn\\
&\xi^{\dagger} \vec \sigma\xi = \hat n,
~~~~~\eta^{\dagger} \vec \sigma\eta= -\hat n, \nn\\
&(\hn \cdot \vec \sigma) ~\xi = \xi,
~~~~~(\hn \cdot \vec \sigma) ~\eta = -\eta,
\label{odbasis}
\eea
and let
\bea
\Phi = \phi_+ \xi + \phi_- \eta, ~~~~~(\phi_+= \xi^{\dagger}\Phi,
~~~\phi_-= \eta^{\dagger} \Phi).
\label{phidecom}
\eea
Now, with the identity
\bea
&\Big[\partial_\mu - \dfrac{g}{2i} \big(C_\mu \hn
+ \dfrac{1}{g} \hn \times \partial_\mu \hn \big)
\cdot \vec \sigma \Big] \xi = 0, \nn\\
&\Big[\partial_\mu + \dfrac{g}{2i} \big(C_\mu \hn
- \dfrac{1}{g} \hn \times \partial_\mu \hn \big)
\cdot \vec \sigma \Big] \eta = 0,
\eea
we find
\bea
\hat D_\mu \Phi = (D_\mu \phi_+) \xi + (D_\mu \phi_-)  \eta,
\label{phidef}
\eea
where
\bea
&D_\mu \phi_+ = (\partial_\mu + \dfrac{g}{2i} {\cal A}_\mu) \phi_+,
~~~D_\mu \phi_- = (\partial_\mu - \dfrac{g}{2i} {\cal A}_\mu) \phi_-, \nn\\
&{\cal A}_\mu = A_\mu + C_\mu, \nn\\
&C_\mu = \dfrac{2i}{g} \xi^{\dagger}\partial_\mu \xi
= - \dfrac{2i}{g} \eta^{\dagger}\partial_\mu \eta. \nn
\eea
From this we can express (\ref{nasclag1}) as
\bea
&{\cal L} = - |D_\mu \phi_+|^2 - |D_\mu \phi_-|^2
+ m^2 (\phi_+^{\dagger}\phi_+ + \phi_-^{\dagger}\phi_-) \nn\\
&- \dfrac{\lambda}{2} (\phi_+^{\dagger}\phi_+ + \phi_-^{\dagger}\phi_-)^2
-\dfrac{1}{4} {\cal F}_{\mu\nu}^2, \nn\\
&{\cal F}_{\mu\nu} = \partial_\mu {\cal A}_\nu 
- \partial_\nu {\cal A}_\mu.
\label{nasclag2}
\eea
This tells that the restricted $SU(2)$ gauge theory
(\ref{nasclag1}) is reduced to
an Abelian gauge theory coupled to oppositely charged
scalar fields $\phi_+$ and $\phi_-$. We emphasize that
this Abelianization is achieved without any gauge fixing.

The Abelianization assures that the non-Abelian theory
is not different from the two-gap Abelian theory.
Indeed with
\begin{equation}
\chi = \left(\begin{array}{rr}
\phi_+\\
\phi_-^*
\end{array}\right),
\end{equation}
we can express the Lagrangian (\ref{nasclag2}) as \cite{cm7}
\bea
&{\cal L} = - |D_\mu \chi|^2 + \mu^2 \chi^{\dagger}\chi
- \dfrac{\lambda}{2} (\chi^{\dagger} \chi)^2
- \dfrac{1}{4} {\cal F}_{\mu \nu}^2, \nn\\
&D_\mu \chi = (\partial_\mu + ig {\cal A}_\mu) \chi,
\label{nalag}
\eea
This is formally identical to the Lagrangian (\ref{sclag}) of two-gap
Abelian superconductor. The only difference is
that here $\phi$ and $A_\mu$ are replaced by $\chi$ and ${\cal A}_\mu$.
This establishes that, with the proper redefinition of field
variables (\ref{phidecom}) and (\ref{phidef}), our non-Abelian
restricted gauge theory of superconductivity can in fact be made
identical to the Abelian gauge theory of two-gap superconductor. This
proves the existence of non-Abelian superconductors
made of the doublet consisting of oppositely charged
condensates. As importantly our analysis tells that
the two-gap Abelian superconductor has a hidden
non-Abelian gauge symmetry because it can be transformed to
the non-Abelian restricted gauge theory. This implies that
the underlying dynamics of two-gap superconductor
is indeed the non-Abelian gauge symmetry.
In the non-Abelian superconductor it is explicit.
But in the two-gap Abelian superconductor it is hidden,
where the full non-Abelian gauge symmetry only
becomes transparent when one embeds the nontrivial topology
properly into the non-Abelian symmetry \cite{cm7}.

Once the equivalence of the two Lagrangians (\ref{sclag})
and (\ref{nasclag1}) is established, it must be evident
that our non-Abelian theory of superconductivity also
allows the non-Abelian vortex which has the non-Abelian 
flux quantization, and thus the non-Abelian
Meissner effect. 

At this point one may ask how realistic is the above theory
of non-Abelian superconductivity. In particular, one may ask 
why the oppositely charged condensates should form a doublet.
To answer this question we notice that $\phi_+$ and $\phi_-$ 
can always be put into a doublet (at least formally). 
So the real question is whether the interaction potential  
can be made $SU(2)$ symmetric or not. 
The answer, of course, depends on materials. In fact
we expect that the symmetry will be broken in real materials.
Nevertheless one may still treat the $SU(2)$ symmetry as an
approximate symmetry. In this case one can adopt the above 
Landau-Ginzburg potential as a simplest potential in the first 
approximation, and treat the symmetry breaking interaction  
perturbatively in two-gap superconductor. In this sense
we believe that the above theory
of non-Abelian superconductivity could be able to describe
the qualitative, if not quantitative, features of two-gap 
superconductor made of oppositely charged condensates.

This implies two things. An immediate implication is 
the existence of the non-Abelian magnetic vortex which has 
non-Abelian flux quantization in ${\rm MgB_2}$. The above discussion 
tells that a two-gap superconductor, regardless of the charge 
content of the doublet condensate, should allow the non-Abelian 
magnetic vortex. The other implication is that the non-Abelian
dynamics could play a crucial role in condensed 
matter physics, in particular in multi-component condensed 
matter. Our result makes it clear that,
implicitly or explicitly, the underlying dynamics of
multi-component condensed matters
can ultimately be related to a non-Abelian dynamics.

{\bf ACKNOWLEDGEMENT}

~~~We thank G. Sterman for the kind hospitality during his
visit at C.N. Yang Institute for Theoretical Physics.
The work is supported in part by the ABRL Program of
Korea Science and Enginnering Foundation (Grant R14-2003-012-01002-0)
and by the BK21 Project of Ministry of Education.

\end{document}